\documentclass[5p,times]{elsarticle}

\usepackage{amsmath}
\usepackage{microtype}

\begin{document}

\title{Multiple phases in sputtered Cr$_2$CoGa films}

\author{Manuel P. Geisler}
\author{Markus Meinert}
\ead{meinert@physik.uni-bielefeld.de}
\author{Jan Schmalhorst}
\author{G\"unter Reiss}
\address{Center for Spinelectronic Materials and Devices, Department of Physics, Bielefeld University, D-33501 Bielefeld, Germany}
\author{Elke Arenholz}
\address{Advanced Light Source, Lawrence Berkeley National Laboratory, CA 94720, USA}

\journal{Journal of Alloys and Compounds}

\date{\today}

\begin{abstract}

By magnetron co-sputtering, thin films of a nominal Cr$_2$CoGa compound were deposited on MgO and MgAl$_2$O$_4$. To achieve crystallisation in the inverse Heusler structure, different heat treatments were tested. Instead of the inverse Heusler structure, we observed phase separation and precipitate formation in dependence on the heat treatment. The main precipitate is Cr$_3$Ga in A15 structure. The remainder forms Co-rich CoGa in the B2 structure and possibly Cr-rich CoCr in the $\sigma$-phase.

\end{abstract}

\maketitle

\section{Introduction}

Theoretical work on Cr$_2$CoGa in inverse Heusler structure predicts it to be a nearly fully compensated half-metallic ferrimagnet as a consequence of the Slater-Pauling rule \cite{Skaftouros2012}. Further calculations indicate a Curie temperature of 1520\,K \cite{Galanakis2011}, which would be the largest among known Heusler compounds \cite{Kubler2007}. Such a material has the potential to be an ideal electrode for spin torque devices \cite{Graf11}. Experimentally, several attempts to synthesise Cr$_2$CoGa in bulk were already performed. According to them, the compound crystallizes in a cubic structure, which could be the inverse Heusler structure X$_a$ \cite{Hakimi2013,Graf2009}. This structure has an fcc lattice with a four atom basis at the internal coordinates $\mathrm{A}=(0,0,0)$, $\mathrm{B}=( \frac{1}{4},\frac{1}{4},\frac{1}{4})$, $\mathrm{C}=( \frac{1}{2},\frac{1}{2},\frac{1}{2})$ and $\mathrm{D}=( \frac{3}{4},\frac{3}{4},\frac{3}{4})$. The four atoms of a compound $X_2YZ$ have the occupation sequence $Y-X-X-Z$ such that inversion symmetry is not present. This structure belongs to space group $F \overline{4} 3 m$ (No. 216).  However, the system appears to be highly susceptible for atomic disorder and a much lower Curie temperature of about 300\,K was found \cite{Qian2011}.

In order to explain this discrepancy, a stability analysis of Cr$_2$CoGa in X$_a$ structure was performed in a previous work \cite{Meinert2013}. It was found that the inverse Heusler phase of this compound is unstable with respect to both constituent elements, binary and ternary decompositions. This instability is indicated by the slightly positive formation energy $\Delta E_0^{\mathrm{Cr_2CoGa}}=0.08$\,eV\,/\,f.u. and negative formation energies for binary phases. The formation of multiple phases is consequently energetically more preferable than the single ternary phase. However, the formation energy of the X$_a$ phase is rather close to zero and it is known that calculated formation energies may differ from actual formation enthalpies at typical synthesis temperatures \cite{Yin13,Grabowski2009}. These differences are typically of the order 0.1\,eV/atom which may enable the existence of Cr$_2$CoGa in inverse Heusler structure as a metastable phase.

In this work we investigate nominal Cr$_2$CoGa thin films synthesised via magnetron co-sputtering on single crystal substrates with the aim to enforce the crystallization by epitaxial constraints.

\section{Methods}
Films of 20\,nm thickness were deposited in a magnetron co-sputtering device. Elemental targets with at least 99.99\% purity were used. After deposition, all thin films were covered with 2\,nm of MgO via electron beam evaporation. X-ray fluorescence was used to adjust the film stoichiometry. MgO and MgAl$_2$O$_4$ (MAO) with [001] cut were tested as substrates with deposition temperatures up to 700$^\circ$C and cooling rates below 30\,K/min. Calculations predict a lattice constant of 5.78\,\AA{} for Cr$_2$CoGa \cite{Meinert2013}, so that lattice mismatches of $-2.9$\,\% (MgO) and $+1.0$\,\% (MgAl$_2$O$_4$) are expected when the usual 45$^\circ$ rotation of the unit cells is taken into account.

Further sample characterization methods were x-ray diffraction (XRD) and reflectivity (XRR), scanning electron microscope (SEM) imaging and energy dispersive x-ray (EDX) mappings. XRD and XRR measurements were performed with a Philips X'Pert Pro diffractometer with a Cu anode, Bragg-Brentano and point-focus collimator optics. Additionally, a Euler cradle was used to search for superstructure reflections. The device used for SEM imaging and EDX mappings is a FEI Helios Nanolab 600 with an EDAX Apollo detector.

Measurements of magnetic circular dichroism (XMCD) and specific x-ray absorption (XAS) were performed at the Advanced Light Source, Berkeley, beamline 6.3.1. The absorption signal was determined in total electron yield and transmission with a photodiode monitoring the substrate luminescence \cite{Kallmayer07}. A magnetic field of $\pm1.8$\,T was applied and the spectra were taken at 80\,K.

Phase decompositions were studied theoretically using the data from the \textit{Materials Project} database \cite{materialsproject,materialsprojectonline}. The formation energy of Cr$_2$CoGa was computed with the \textsc{elk} full-potential linearised augmented-plane wave (FP-LAPW) code \cite{elk}. Both the \textit{Materials Project} and our total energy calculations are based on the Perdew-Burke-Ernzerhof (PBE) functional \cite{pbe}. Crystal structure visualizations were generated with VESTA \cite{vesta}.

\begin{figure}[t]
   \centering
\includegraphics[width=8.6cm]{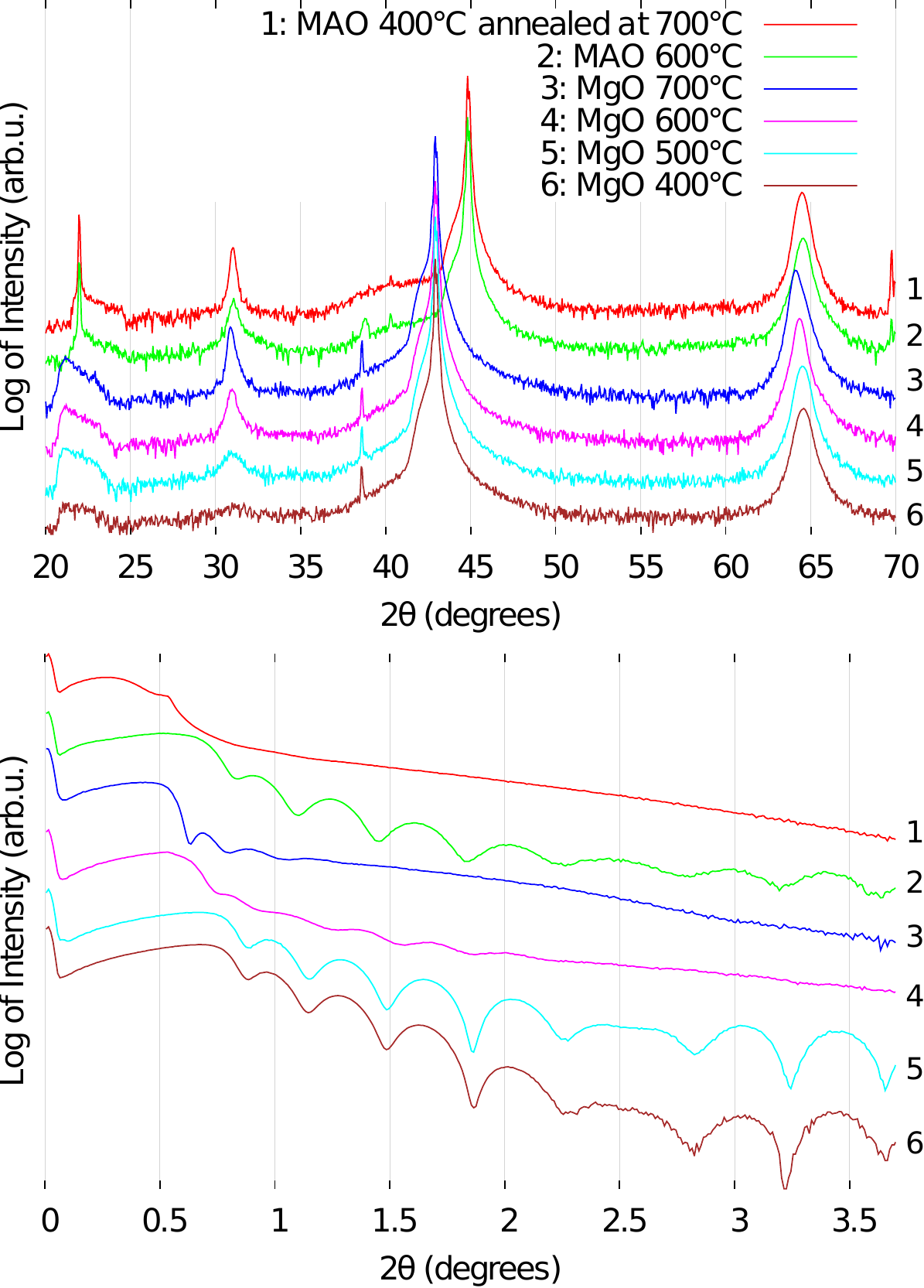}
   \caption[XRD and XRR measurements of Cr2CoGa]{XRD (top) and XRR (bottom) measurements of six different Cr$_2$CoGa samples. All sharp peaks and broad peaks below $2\theta = 25^\circ$ originate from the substrates.}
\label{pic:Cr2CoGaXRD}
\end{figure}

\section{Results}

Figure~\ref{pic:Cr2CoGaXRD} shows results of XRD and XRR measurements of six different samples deposited under various conditions. The XRR spectra indicate that the films get rough at deposition or annealing temperatures of 600$^\circ$C and more. The main diffraction peaks at 30.9$^\circ$ and  64.4$^\circ$ have distinct deposition temperature dependences. While the latter reflection varies not too much, the first reflection gets more intense and narrower with increasing deposition temperature. Assuming the observed main reflections belong to the $X_a$ phase, we define the order parameter
\begin{equation}
S = \sqrt{\frac{I_{(002)}^\mathrm{exp.} / I_{(004)}^\mathrm{exp.}}{I_{(002)}^\mathrm{theo.} / I_{(004)}^\mathrm{theo.}}},
\end{equation}
which is sensitive to disorder between A/B sites, A/D sites, and C/D sites, whereas it is insensitive to disorder between A/C sites, B/D, and B/C sites. In the case of Heusler compounds, it is equivalent to S$_\mathrm{B2}$ \cite{sb2} and is related to the number of $X$ atoms on $Y$ or $Z$ sites. With the above definition we found $S_{\mathrm{1}}=1.07$, $S_{\mathrm{2}}=1.21$, $S_{\mathrm{3}}=1.11$ and $_{\mathrm{4}}=1.04$, where the subscript numbers refer to the numbers in Figure~\ref{pic:Cr2CoGaXRD}. For A/B disorder or strong C/D disorder we can have $1 < S < 1.092$. In both cases, the structure changes to the regular Heusler type, which is however energetically highly unfavorable ($+0.65$\,eV\,/\,f.u. with respect to the inverse type \cite{Galanakis2011}). Further, the large $S_2$ cannot be explained this way. Therefore we conclude that the two peaks do not originate from the X$_a$ phase nor from the regular Heusler phase. For traces 5 and 6 we have $S_{\mathrm{5,6}} \ll 1$, so a possible superstructure is only weakly present. At high deposition temperatures, a broad reflection around $2\theta = 39^\circ$ is observed, particularly on MAO. We have searched for the $(111)$ or $(11\bar{1})$ superstructure reflections of a possible inverse Heusler phase to which the two main reflections may belong by pole figure scans around $2\theta = 26.7^\circ$. The reflections could not be observed, which excludes the formation of the Heusler or inverse Heusler structure. 

\begin{figure}[t]
  \centering
\includegraphics[width=8.6cm]{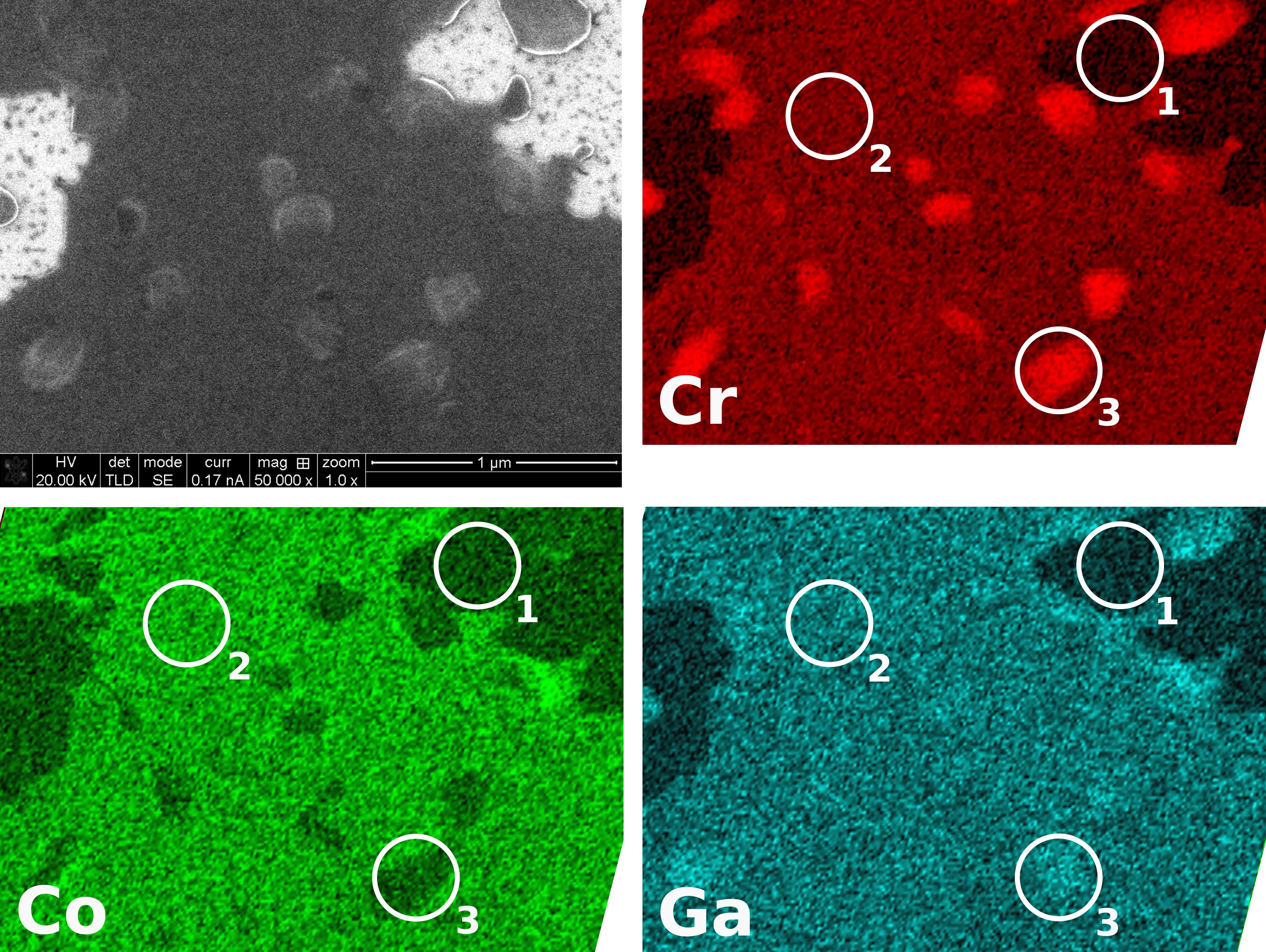}
   \caption[Mapping image of Cr$_2$CoGa]{SEM and EDX mapping images of a nominal Cr$_2$CoGa sample on MAO deposited at 600$^\circ$C.}
\label{pic:Cr2CoGamapping}
\end{figure}

\begin{figure}[t]
  \centering
\includegraphics[width=8.6cm]{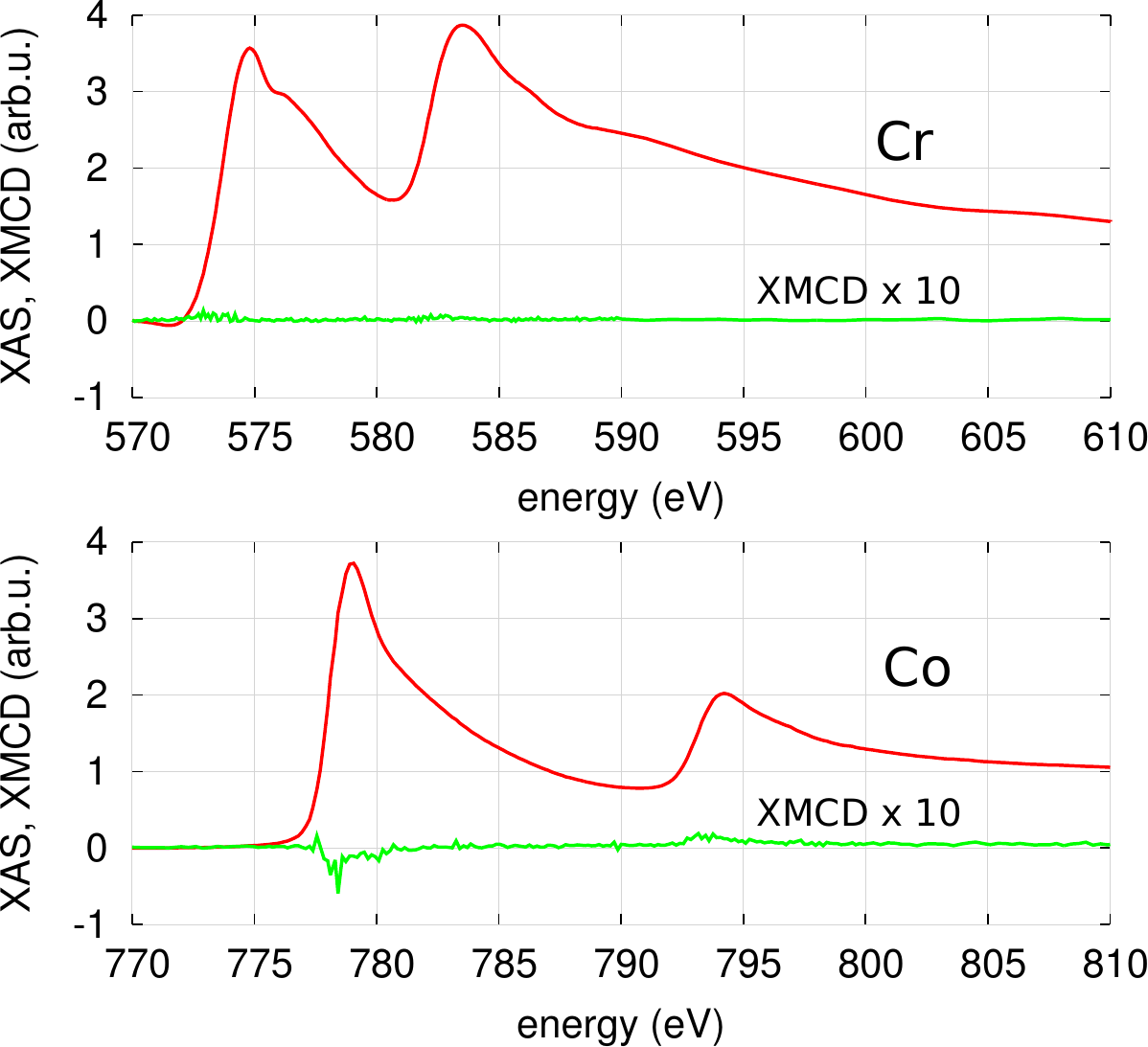}
   \caption[XAS and XMCD spectra of Cr$_2$CoGa]{XAS and XMCD spectra on Cr and Co $L_{3,2}$ edges of a nominal Cr$_2$CoGa sample deposited at 600$^\circ$C on MgO. The spectra were taken in total electron yield at 80\,K.}
\label{pic:XAS}
\end{figure}

To investigate the deposition temperature dependent film roughness, SEM images and EDX mappings were taken. Figure~\ref{pic:Cr2CoGamapping} shows the images taken on a  sample deposited at 600$^\circ$C on MAO. As indicated, the mapping reveals the existence of three rather different regions of the sample. Region 1 and areas similar to it do not show a film, giving rise to the charging in the SEM image. Most parts of the mapped area look like region 2, where all sputtered elements can be found. Region 3 is an example of grains of Cr and Ga, which are essentially free of Co. Other samples look quite similar to the sample shown here.

Figure~\ref{pic:XAS} exemplarily shows the results of XMCD measurements of a sample deposited at 600$^\circ$C on MgO. The weak XMCD signal points to either essentially nonmagnetic Cr and Co, or to nearly compensated ferrimagnetism for each element individually. The shoulders visible in the Cr XAS can be attributed to oxidation; they are more pronounced in rougher samples, where the protective MgO layer is not thick enough to prevent the oxidation. The very weak XMCD is present both in surface sensitive electron yield spectra and in the bulk sensitive transmission spectra. The Co XAS is nearly free of additional structures, in contrast to typical XAS line shapes of Co in Heusler and inverse Heusler compounds \cite{Telling08, Meinert11}. All samples that we have investigated had similarly weak XMCD signals.

\begin{table*}[t]
\begin{tabular}{l r}
\hline\hline
reaction											& $\Delta E_0$ (eV/f.u.)\\\hline
a) 2(2Cr + Co + Ga)		$\rightarrow$     Co$_2$CrGa + Cr$_3$Ga  	& -0.61\\
b) 2Cr + Co + Ga    	$\rightarrow$     GaCo + 2Cr			        & -0.56\\
c) 2(2Cr + Co + Ga)		$\rightarrow$     Cr$_3$Ga + GaCo + Co + Cr 	& -0.38\\
d) 3(2Cr + Co + Ga)		$\rightarrow$     Ga$_3$Co + 6Cr + 2Co      	& -0.35\\
e) 2(2Cr + Co + Ga)		$\rightarrow$     2Co(Ga$_{0.5}$Cr$_{0.5}$) + Cr$_3$Ga  & -0.18\\
\textbf{f) 2Cr + Co + Ga}		$\rightarrow$	  \textbf{Cr$_2$CoGa} & \textbf{0.08} \\
\hline\hline
\end{tabular}
\caption[Decompositions of Cr$_2$CoGa]{\label{Cr2CoGabin} Most favorable reactions forming binary and ternary phases and their reaction energies according to the \textit{Materials Project} database for $\mathrm{Cr_2CoGa}$.}
\end{table*}

\begin{figure}[t]
  \centering
\includegraphics[width=8.6cm]{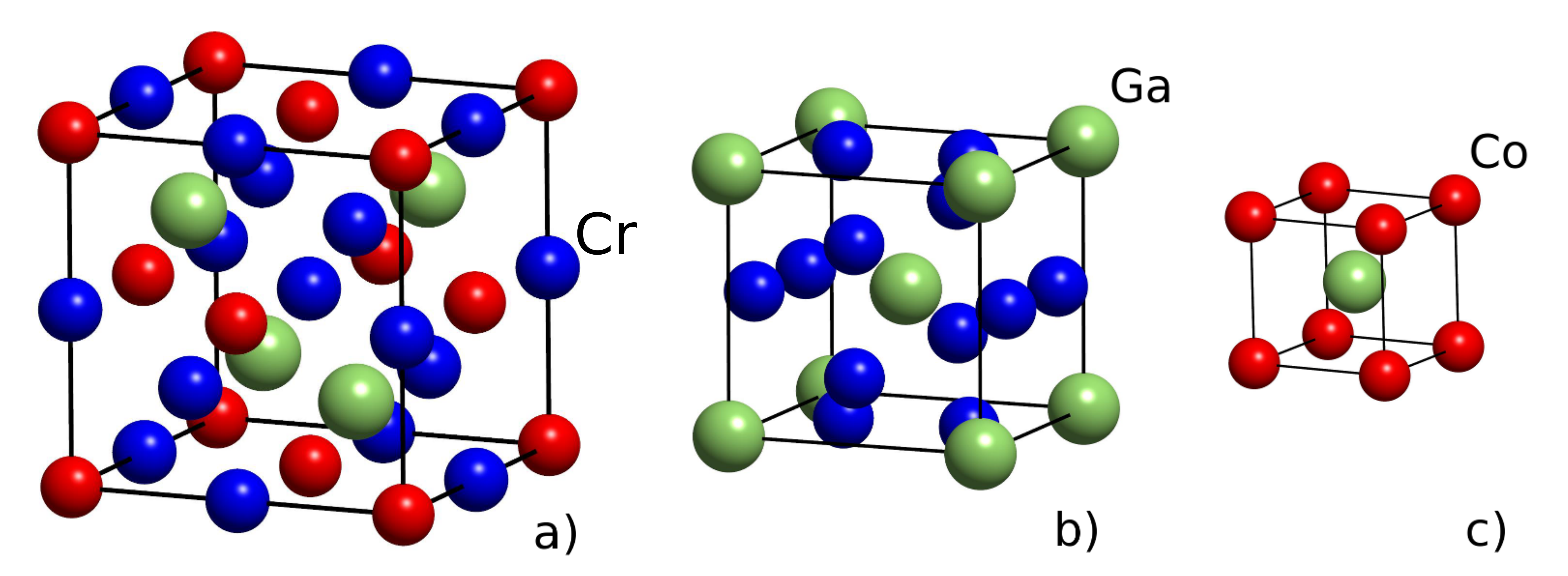}
   \caption{a): Ideal inverse Heusler structure of Cr$_2$CoGa. b): Cr$_3$Ga in A15 structure. c): CoGa in B2 structure.}
\label{pic:structures}
\end{figure}

Taking together the experimental findings, we attempt to identify the phases into which the Cr$_2$CoGa films separate with guidance from the computed formation energies in Table~\ref{Cr2CoGabin}, the binary phase diagrams \cite{Okamoto05, Okamoto07, Ishida90} and the high-temperature ternary phase diagrams of Co-Cr-Ga \cite{Kobayashi05}. The Co-poor precipitates are most likely binary Cr-Ga compounds. In the Cr-Ga system four intermetallic phases are known, of which CrGa decomposes below 736$^\circ$C. The three stable phases are Cr$_3$Ga with A15 structure (space group $Pm\bar{3}n$, cf. Fig \ref{pic:structures} b)), CrGa$_4$ with PtHg$_4$ structure (space group $Im\bar{3}m$) and Cr$_5$Ga$_6$ with unknown structure. All these phases are either para- or diamagnetic. According to the EDX mapping, the precipitates are rich in Cr, so it is likely that they are mainly Cr$_3$Ga.  The broad reflection observed in the XRD spectra around $2\theta = 39^\circ$ can be explained as the (002) reflection of this phase with $a \approx 4.6$\,\AA. In view of the broad reflection, these precipitates have to be polycrystalline with a crystallite size around 2\,nm. The formation of this phase is also very likely in terms of the formation energy, see Table \ref{Cr2CoGabin}; the remainder of the film could crystallize as Co$_2$CrGa in the regular Heusler structure, which is known to exist \cite{Kobayashi05}, and is seemingly energetically favorable. This could account for the main x-ray reflections around $2\theta \approx 31^\circ$ and $2\theta \approx 64^\circ$, but the corresponding (111) reflection around  $2\theta = 26.7^\circ$ was not observed. Assuming a B2-type (cf. Figure \ref{pic:structures} c)) disordered phase like Co(Ga$_{x}$Cr$_{1-x}$) with $x=0.5$, which is energetically less favorable, would explain the missing (111) reflection. However, it is also calculated to be a ferromagnet with 1.67$\mu_\mathrm{B}$ on Cr and 0.64$\mu_\mathrm{B}$ on Co, at odds with the XMCD results. Thus, we assume that either the deposition temperature was insufficient to form Co$_2$CrGa or it decomposes during the slow cooling. It is known that various phases occur in the Co-Cr-Ga system near the Co$_2$CrGa composition below $800^\circ$C, namely hcp-Co, B2 CoGa, and D8$_\mathrm{b}$ CoCr (also known as $\sigma$-CoCr) \cite{Kobayashi05}. Only at temperatures higher than our deposition temperature, significant amounts of Cr will dissolve in B2 CoGa. Thus, CoGa with $a=2.88\,\AA{}$ and a strong (001) texture due to the epitaxial matching with the substrate is the best candidate to explain the observed XRD spectra. The two main peaks can be indexed as the (001) and (002) reflections of this phase and no reflections below $2\theta = 31^\circ$ would be present. This compound is paramagnetic and becomes weakly ferromagnetic at increased Co content \cite{Booth70}. The Co-rich B2 phase is stable up to about 63at.\% Co at our deposition or annealing conditions. At this composition, a Curie temperature around 350\,K and a magnetization of $0.8\,\mu_\mathrm{B}$ per unit cell may be expected from extrapolation of solid-solution experiments:
\begin{align}
T_\mathrm{C} &= 2.33 x^2 \,\,\mathrm{K/(at.\%)^2},\\
m &= 0.31 x^2 \,\,\mathrm{emu/g/(at.\%)^2},
\end{align}
with $x$ being the excess amount of Co ($50 + x$) given in at.\% \cite{Booth70}. Quenching from high temperature would freeze this state and the above values are in agreement with earlier experiments on quenched bulk material \cite{Hakimi2013, Qian2011}. In our case, the magnetic moment of Co at 80\,K is virtually zero, so that this extrapolation allows us to conclude that there is less than 6\% excess Co, as one expects from the binary phase diagram for slowly cooled CoGa \cite{Okamoto05}. The expected (001)/(002) peak intensity ratio matches the observed values when about 15\% disorder between Co and Ga sites is taken into account. More disorder strongly diminishes the (001) reflection as in traces 5 and 6 of Figure \ref{pic:Cr2CoGaXRD}, which represent samples with lower deposition temperatures.
    
The remainder of the material could form a Cr-rich $\sigma$-phase \cite{Kobayashi05}, which would be weakly magnetic \cite{Pavlu10}. If these crystallites remained sufficiently small, they could not be detected in the EDX mapping or with XRD. The reaction
\begin{equation}
\mathrm{2(2Cr + Co + Ga) \rightarrow Cr_3Ga + Co_{1+x}Ga + Co_{1-x}Cr}
\end{equation}
is thus our final conjecture for the decomposition of nominal Cr$_2$CoGa, which explains the XRD, EDX mapping, and element specific magnetization data of our films.

\section{Conclusions}
We were able to show experimentally that Cr$_2$CoGa thin films deposited via magnetron sputtering do not form the desired half-metallic and ferrimagnetic X$_a$ structure. Instead we find a segregation into CoGa with B2 and Cr$_3$Ga with A15 structure. Possibly also the $\sigma$-phase of CoCr is formed. The instability of the Cr$_2$CoGa compound makes this material unsuitable for the proposed applications in spintronics.

\section*{Acknowledgments}
This work was supported by the European Commission via the Collaborative Project HARFIR (Project No. 604398). Further financial support by the Deutsche Forschungsgemeinschaft (DFG) is gratefully acknowledged. We thank Dr. Karsten Rott for help with EDX mappings and SEM images. The Advanced Light Source is supported by the Director, Office of Science, Office of Basic Energy Sciences, of the U.S. Department of Energy under Contract No. DE-AC02-05CH11231.

\end{document}